\documentclass{amsart}

\newtheorem{theorem}{Theorem}
\theoremstyle{remark}
\newtheorem{remark}{Remark}

\begin{document}

\title[Microscopic Derivation of the Ginzburg--Landau Model]{MICROSCOPIC DERIVATION OF\\ THE GINZBURG--LANDAU MODEL}

\thanks{Contribution to the proceedings of ICMP12, Aalborg, Denmark, August 6--11, 2012. \\ \copyright\, 2012 by
  the authors. This paper may be reproduced, in its entirety, for
  non-commercial purposes.}

\author{R.L. Frank}

\address{Department of Mathematics, Princeton University\\ Princeton, NJ
  08544, USA\\ Email: rlfrank@math.princeton.edu}

\author{C. Hainzl}

\address{Mathematisches Institut, Universit\"at T\"ubingen\\ Auf der
  Morgenstelle 10, 72076 T\"ubingen, Germany\\ Email: christian.hainzl@uni-tuebingen.de}

\author{R. Seiringer}

\address{Department of Mathematics and Statistics, McGill
  University\\ 805 Sherbrooke Street West, Montreal, QC H3A 2K6,
  Canada\\ Email: robert.seiringer@mcgill.ca}

\author{J.P. Solovej}

\address{Department of Mathematics, University of Copenhagen\\
  Universitetsparken 5, DK-2100 Copenhagen, Denmark\\
Email: solovej@math.ku.dk}

\begin{abstract}
  We present a summary of our recent rigorous derivation of the
  celebrated Ginzburg--Landau (GL) theory, starting from the
  microscopic Bardeen--Cooper--Schrieffer (BCS) model. Close to the
  critical temperature, GL arises as an effective theory on the
  macroscopic scale. The relevant scaling limit is semiclassical in
  nature, and semiclassical analysis, with minimal regularity
  assumptions, plays an important part in our proof.
\end{abstract}

\maketitle

\section{Introduction}\label{aba:sec1}

The purpose of this paper is to describe how the celebrated
Ginzburg--Landau (GL) model of superconductivity \cite{GL} arises as
an asymptotic limit of the microscopic Bardeen--Cooper--Schrieffer
(BCS) model \cite{BCS}. The relevant asymptotic limit may be seen as a
semiclassical limit, and one of the main difficulties involved in the
proof is to derive a semiclassical expansion with minimal regularity
assumptions. The present article represents a summary of our recent
work in \cite{FHSS} (see also \cite{FHSS2} for a simplified model in
one dimension) and we shall refer to \cite{FHSS} for all technical
details.

We emphasize that it is not rigorously understood how the BCS model
approximates the underlying many-body quantum system. We will
formulate the BCS model as a variational problem, but only
heuristically discuss its relation to quantum mechanics.

\section{The Ginzburg--Landau Model}\label{sec:gl}

The Ginzburg--Landau model is a phenomenological model for
superconducting materials. Imagine a sample of such a material
occupying a three-dimensional box $\Lambda$.\footnote{For simplicity,
  we restrict our attention to three dimensions here, but a similar
  analysis applies in one and two dimensions as well.} If $W$ denotes
a scalar external potential, and ${\bf A}$ a magnetic vector
potential, the Ginzburg--Landau functional is given as
\begin{eqnarray}\nonumber
\mathcal{E}^{\rm GL}(\psi) = &\displaystyle\int_{{\Lambda}}& \biggl(
|(-i\nabla + 2{\bf A}(x))\psi(x)|^2
   + \lambda_1 {W}(x) |\psi(x)|^2 \\&&-\lambda_2 |\psi(x)|^2 + \lambda_3 |\psi(x)|^4  \biggl) dx\,, \label{def:GL}
\end{eqnarray}
where $\psi\in H^1(\Lambda)$. Here, the $\lambda_i$ are real
parameters, with $\lambda_3>0$. Note the factor $2$ in front of the
${\bf A}$-field, which is reminiscent of the fact that $\psi$
describes pairs of particles. The microscopic model discussed in the
next section does not have such a factor $2$, as it describes single
particles.

By simple rescaling, we
could take $\lambda_3=2 |\lambda_2|$ if $\lambda_2\neq 0$. If $\lambda_2>0$, one could
then complete the square in the second line of (\ref{def:GL}) and
write it as $\lambda_3 ( 1 - |\psi(x)|^2 )^2 - \lambda_3$ instead.\footnote{This
  is, in fact, the convention used in \cite{FHSS}.} Since $\lambda_2$
can have either sign, however, we prefer to use the more general
formulation in (\ref{def:GL}) here. We shall derive formulas
for the coefficients $\lambda_i$ from the BCS model below.

The function $\psi$ is interpreted as the order parameter of the
system. In the absence of external fields, i.e., for $W=0$ and ${\bf
  A}={\bf 0}$, the minimum of (\ref{def:GL}) is attained at
$|\psi(x)|^2= \lambda_2/(2\lambda_3)$ for $\lambda_2>0$, or at $\psi=0$ for
$\lambda_2\leq 0$, respectively.

Our main concern here will be the relation of (\ref{def:GL}) with the
Bardeen--Cooper--Schrieffer theory of superconductivity, which
we describe in the next section.  We will derive GL from BCS in an
appropriate limit, where the temperature is close to the critical one,
and the external fields ${\bf A}$ and $W$ are suitably small and slowly
varying on a microscopic scale.

We note that there is a considerable literature \cite{FH,Serfaty} concerning functionals of the type (\ref{def:GL}) and their minimizers, usually with an additional term added corresponding to the magnetic field energy. Such a term plays no role here since ${\bf A}$ is considered a fixed, external field.

\section{The BCS Energy Functional}

The BCS model is based on the microscopic many-particle Hamiltonian
describing the system under consideration. Consider a gas of spin $1/2$
fermions confined to the box $\Lambda$. With $\mu$ denoting the
chemical potential, the Hamiltonian is given by
\begin{equation}\label{def:ham}
H =\sum_j \left( \left(- i \nabla_j + {\bf A}(x_j) \right)^2  - \mu + W(x_j) \right) + \sum_{i<j} V(|x_i -x_j|) \,.
\end{equation}
Here, $V$ is the two-particle interaction potential, which is taken to
be local, i.e., it is a multiplication operator. In the original BCS
Hamiltonian \cite{BCS} it is replaced by an effective, non-local
interaction which results from integrating out the phonon
variables. For simplicity, we prefer to work with a local interaction
of the form (\ref{def:ham}) here.\footnote{It is not necessary to assume $V$ to be radial, as done here, but the formulas simplify somewhat in this case.}

The BCS model can be obtained from (\ref{def:ham}) by an approximation
procedure (see Appendix~A in \cite{HHSS}). One first
restricts the states of the system to quasi-free states on Fock
space. These states are completely determined by two quantities, the
one-particle reduced density matrix\footnote{We suppress the spin dependence in the notation for simplicity.}
\begin{equation}
\gamma(x,y) = \left\langle a^\dagger(x) a(y) \right \rangle
\end{equation}
and  the Cooper-pair wave function
\begin{equation}
\alpha(x,y) = \left\langle a(x) a(y) \right \rangle \,.
\end{equation}
One thus obtains an expression for the energy of such states solely
in terms of $\gamma$ and $\alpha$. One further ignores the direct and
exchange term in this expression, and arrives at the following BCS
energy functional. With $T\geq 0$ denoting the temperature of the
system,
\begin{eqnarray}\nonumber
\mathcal{F}(\Gamma)&=& {\rm Tr\,} \left[ \left(
    \left(-i \nabla + {\bf A}(x) \right)^2 -\mu + W(x)\right) \gamma \right]
- T\, S(\Gamma)\\&& + \int_{\Lambda\times\Lambda}
  V(|x-y|) |\alpha(x,y)|^2 \, {dx \, dy} \,. \label{def:BCS}
\end{eqnarray}
The BCS states can be expressed as $2\times 2$ block matrices
\begin{equation}
\Gamma=\left(\begin{array}{cc}\gamma&\alpha\\\overline{\alpha}&1-
\overline{\gamma}
\end{array}\right)
\end{equation}
satisfying the constraint $0 \leq \Gamma \leq 1$. The entropy $S(\Gamma)$ in (\ref{def:BCS}) takes the usual form  $S(\Gamma)= - {\rm Tr\,} \Gamma \ln \Gamma$.

One of the main questions concerning the model (\ref{def:BCS}) is
whether a minimizing state $\Gamma$ has $\alpha$ identically zero or
not, i.e., whether or not Cooper pairs exist. The non-vanishing of
$\alpha$ implies a superconducting (or superfluid, depending on the
context) behavior of the system.

\subsection{BCS Energy in the Translation-Invariant Case}\label{ss:ti}

It was shown in \cite{HHSS} that in the absence of external
fields, i.e., for ${\bf A}={\bf 0}$ and $W=0$, there exists a critical
temperature $T_c\geq 0$ such that
\begin{itemize}
\item $T\geq T_c$: Minimizer is \emph{normal}, i.e., $\alpha=0$ and $\gamma= ( 1 + \exp((-\nabla^2-\mu)/T))^{-1}$  
\item $T<T_c$: Minimizer has $\alpha\ne0$. 
\end{itemize}
This critical temperature may be characterized by the operator
\begin{equation}\label{def:KT}
K_{T_c}(-\nabla^2-\mu)+V(|x|)\ ,  \qquad K_T(\eta)=\frac{\eta}{\tanh(\eta/2T)}
\end{equation}
having zero as the lowest eigenvalue. Note that the spectrum of
$K_T(-\nabla^2-\mu)$ equals $[2T,\infty)$, hence $0$ is necessarily an
isolated eigenvalue of (\ref{def:KT}) if $T_c>0$.

This linear criterion on the critical temperature was used to derive
precise asymptotics of $T_c$ for weak coupling and/or low density, see
\cite{bcs1}--\cite{bcs3}.

{\em In the following, we shall assume} that $V$ is such that $T_c>0$, and
that the eigenfunction $\alpha_0$ corresponding to the zero eigenvalue
of (\ref{def:KT}) is unique. The potential $V$ also has to be sufficiently regular to
be form-bounded with respect to the Laplacian, and should decay at
infinity.

\subsection{Microscopic vs. Macroscopic Scales}

Let us introduce a small parameter $h>0$, describing the ratio between
the microscopic and macroscopic length scales. The external fields
${\bf A}$ and $W$ occurring in the GL functional vary on the macroscopic
scale, i.e., on the scale of the box $\Lambda$.  The particle
interaction $V$, on the other hand, varies on the microscopic
scale. To take this into account, we replace the external fields ${\bf
  A}(x)$ and $W(x)$ in the BCS functional (\ref{def:BCS}) by
\begin{equation}
\widetilde {\bf A}(x)  = h {\bf A}(hx)\ , \quad \widetilde W(x)  = h^2 W (hx)
\end{equation}
and define the BCS functional on a rescaled box $\widetilde \Lambda =
h^{-1}\Lambda$. Here, $x$ is the microscopic variable, while $\widetilde x = h x$ is the macroscopic variable. 

We find it more convenient to express the BCS functional in
macroscopic variables, and shall henceforth drop the
$\widetilde{\phantom{x}}$'s. The resulting rescaled BCS functional is
\begin{eqnarray}\nonumber
  \mathcal{F}({\Gamma})&=& {\rm Tr\,} \left[ \left(
      \left(-ih \nabla + h{\bf A}(x) \right)^2 -\mu + h^2W(x)\right) \gamma \right] - T\, S(\Gamma)
  \\&& + \int_{\Lambda\times\Lambda}
  V(h^{-1}|x-y|) |\alpha(x,y)|^2 \, {dx \, dy} \,, \label{def:rBCS}
\end{eqnarray}
and it is defined on the $h$-independent macroscopic volume
$\Lambda$. Note the semiclassical nature of the appearance of $h$ in
the various terms in (\ref{def:rBCS}). For small $h$ the energy is of order $h^{-3}$. 

In order for (\ref{def:rBCS}) to be well-defined, suitable boundary
conditions have to be imposed on the boundary of $\Lambda$. In the
limit $h\to 0$, these are not relevant in the bulk of the sample. In
order to avoid technical problems related to these boundary condition,
we chose in \cite{FHSS} to work with an infinite system
instead, which is assumed to be periodic, and all energies are
calculated per unit volume. Consequently, also the functions $\psi$ in
the GL functional (\ref{def:GL}) have to be periodic.

\section{Main Results}

We shall choose the temperature $T$ in (\ref{def:rBCS}) to be close to the critical temperature defined in Subsection~\ref{ss:ti}. More precisely, we take  
\begin{equation}\label{def:D}
T=T_c(1-Dh^2)
\end{equation}
for some $h$-independent parameter $D\in \mathbb{R}$.\footnote{The
  results in \cite{FHSS} were stated for $D>0$, but their
  proof is equally valid for $D\leq 0$.}

Our main result concerns the asymptotic behavior of $\inf_\Gamma \mathcal{F}(\Gamma)$ and the corresponding minimizers as $h\to 0$.

\begin{theorem}\label{main:thm}
There exists a $\lambda_0>0$ and parameters $\lambda_1$, $\lambda_2$ and $\lambda_3$ in the GL functional
such that 
\begin{equation}
\inf_{\Gamma}\mathcal{F}(\Gamma)=\mathcal{F}(\Gamma_0) + \lambda_0 h \inf_\psi\mathcal{E}^{\rm GL}(\psi) +o(h) 
\end{equation}
as $h\to0$, where $\Gamma_0$ is the normal state (i.e., the minimizer in the absence of $V$).

Moreover, if $\Gamma$ is a state such that $\mathcal{F}(\Gamma)\leq \mathcal{F}(\Gamma_0) + \lambda_0 h  \inf_\psi\mathcal{E}^{\rm GL}(\psi) + o(h)$
then the corresponding Cooper pair wave function $\alpha$ satisfies
\begin{equation}
\|\alpha-\alpha_{\rm GL}\|^2_{L^2}\leq o(h)\|\alpha_{\rm GL}\|^2_{L^2}=o(h)h^{-1}
\end{equation}
where
\begin{equation}\label{def:agl}
\alpha_{\rm GL}(x,y) = h^{-2}\psi_0\left(\frac{x+y}{2}\right)
    \alpha_0\left(\frac{x-y}{h}\right) = \hbox{Op}(h\psi_0(x)\widehat\alpha_0(p))
\end{equation}
 and 
$\mathcal{E}^{\rm GL}(\psi_0) \leq \inf_\psi\mathcal{E}(\psi)+ o(1)$.
\end{theorem}

Theorem~\ref{main:thm} represents a rigorous derivation of
Ginzburg--Landau theory. Starting from the BCS model, GL arises as an
effective theory on the macroscopic scale, in the presence of weak and
slowly varying external fields, and for temperatures close to the
critical one.

\begin{remark}
As mentioned above, $\mathcal{F}(\Gamma_0)$ is $O(h^{-3})$, hence the
GL functional arises as an $O(h^4)$ correction to the main term.
\end{remark}

\begin{remark}
In (\ref{def:agl}), $Op$ denotes Weyl
quantization. Theorem~\ref{main:thm} demonstrates the role of the
function $\psi$ in the GL model: It describes the center-of-mass motion of the
Cooper pair wavefunction, which close to the critical temperature
equals (\ref{def:agl}) to leading order in $h$.
\end{remark}

\begin{remark}
  The method of our analysis can also be used to show that the GL
  model predicts the correct change in critical temperature in the BCS
  theory due to the external field \cite{FHSSnew}. As discussed in
  the next subsection, the parameter $\lambda_2$ is proportional to
  the difference between the critical and the actual temperature (more precisely, to $D$ in (\ref{def:D})), and
  the mentioned shift corresponds to the largest $\lambda_2$ with the
  property that $\psi=0$ is a GL minimizer. This can have either sign,
  depending on the external fields ${\bf A}$ and $W$.
\end{remark}

\begin{remark}
  A similar analysis can be used at $T = 0$ to study the low-density
  limit of the BCS model. In this limit, one obtains a Bose-Einstein
  condensate of fermion pairs, described by the Gross-Pitaevskii
  equation \cite{HSnew,Hnew}.
\end{remark}

\subsection{The Coefficients $\lambda_i$}

The coefficients $\lambda_0$, $\lambda_1$, $\lambda_2$ and $\lambda_3$
in Theorem~\ref{main:thm} can be explicitly calculated. They are all
expressed in terms of the eigenfunction $\alpha_0$ corresponding to
the zero eigenvalue of (\ref{def:KT}), the critical temperature $T_c$ and the coefficient $D$ in
(\ref{def:D}).

Specifically, if we denote by $t$ the Fourier transform of $2 K_{T_c}\alpha_0$, we have\footnote{In the published version, there is a misprint in Eq. (\ref{lambda0}): the factor $\frac 23$ in front of the second term in the integrand is wrongly written as $2$.}
\begin{equation}\label{lambda0}
\lambda_0  =  \frac{1}{16 T_c^2 }   \int_{\mathbb{R}^3} t(q)^2 \left(  g_1(\beta_c(q^2-\mu)) + \tfrac 23 \beta_c q^2 \, g_2(\beta_c(q^2-\mu)) \right) \frac{dq}{(2\pi)^3}\,,
\end{equation}
\begin{equation}
\lambda_1 = \lambda_0^{-1}   \frac  {1}{4 T_c^2}  \int_{\mathbb{R}^3} t(q)^2 \, g_1(\beta_c(q^2-\mu)) \, \frac{dq}{(2\pi)^3}  \,,
\end{equation}
\begin{equation}
\lambda_2 = \lambda_0^{-1} \frac{D}{8 T_c} \int_{\mathbb{R}^3}  {t(q)^2} \cosh^{-2} \left( \frac {\beta_c}{2}(q^2 -\mu) \right) \frac{dq}{(2\pi)^3}
\end{equation}
and
\begin{equation}\label{def:b3}
 \lambda_3 =   \lambda_0^{-1} \frac {1} {16 T_c^2}  \int_{\mathbb{R}^3} t(q)^4 \, \frac{g_1(\beta_c(q^2-\mu))}{q^2-\mu}\,\frac{dq}{(2\pi)^3} \,.
\end{equation}
Here $\beta_c= T_c^{-1}$, and $g_1$ and $g_2$ denote the functions
\begin{equation}
g_1(z) = \frac{ e^{2 z} - 2 z e^{z}-1}{z^2 (1+e^{z})^2} \quad \text{and} \quad 
g_2(z) = \frac{2 e^{z} \left( e^{ z}-1\right)}{z
  \left(e^{z}+1\right)^3} \,,
\end{equation}
respectively. 
One can show \cite{FHSS} that $\lambda_0>0$. Note that $g_1(z)/z > 0$,
hence also $\lambda_3>0$. The coefficient $\lambda_2$ is proportional
to $D$, defined in (\ref{def:D}).

As mentioned in Section~\ref{sec:gl}, the terms in the second line of
the GL functional (\ref{def:GL}) are often written as $\frac{\kappa^2}{2}
(1-|\psi(x)|^2)^2$ instead, with a suitable coupling constant
$\kappa>0$. In our notation, $\kappa$ corresponds to 
\begin{equation}
\kappa = \sqrt{\lambda_2}
\end{equation} (in case $D > 0$, i.e., $T<T_c$).

Note that the normalization of $\alpha_0$ is irrelevant. If we
multiply $\alpha_0$ by a factor $\lambda>0$, then $\lambda_0$ and
$\lambda_3$ get multiplied by $\lambda^2$, while $\lambda_1$ and
$\lambda_2$ stay the same. Hence the GL minimizer $\psi_0$ gets
multiplied by $\lambda^{-1}$, leaving both $\lambda_0 \mathcal{E}^{\rm
  GL}(\psi_0)$ and the product $\psi_0 \alpha_0$ unchanged. In particular,
(\ref{def:agl}) is independent of the normalization of $\alpha_0$.

\section{Sketch of Proof}

In the following, we present a very brief sketch of the main ideas in the proof of Theorem~\ref{main:thm}. The actual proof is rather lengthy and we have to refer to \cite{FHSS} for details.

The starting point is the identity
\begin{align}\nonumber
\mathcal{F}(\Gamma) - \mathcal{F}(\Gamma_0) & = - \frac T2 {\rm Tr\,} \left[ \ln\left(1+e^{-H_\Delta/T}\right) - \ln\left(1+e^{-H_0/T} \right) \right] \\ \nonumber & \quad - \int V(|x-y|/h)|\alpha_{\rm GL}(x,y)|^2 dxdy
\\ & \quad + \frac T 2 \mathcal{H}(\Gamma,\Gamma_\Delta) + \int ￼V (|x - y|/h)\left|\alpha_{\rm GL}(x, y) - \alpha(x, y)\right|^2 \, dx\, dy  \label{fi}
\end{align}
where $\Gamma_\Delta = (1+e^{H_\Delta/T})^{-1}$, 
\begin{equation}
H_\Delta=\left(\begin{array}{cc}\mathfrak{h}&\Delta\\
\overline{\Delta}&-\overline{\mathfrak{h}}
\end{array}\right)\ ,\quad 
\mathfrak{h}=\left(-ih \nabla + h{\bf A}(x) \right)^2 -\mu +
    h^2 W(x)
\end{equation}
and $\Delta$ denotes the operator with integral kernel
\begin{equation}
\Delta(x,y)=2V(|x-y|/h)\alpha_{\rm
  GL}(x,y)=2h\hbox{Op}(\psi_0(x)(\widehat{\alpha_0 V})(p)) \,.
\end{equation}
Moreover, $\mathcal{H}(\Gamma,\Gamma_\Delta)$ denote the relative entropy
\begin{equation}
\mathcal{H}(\Gamma,\Gamma_\Delta) = {\rm Tr\,} \left[ \Gamma\left( \ln \Gamma - \ln \Gamma_\Delta \right) + \left(1-\Gamma\right) \left( \ln \left(1-\Gamma\right) - \ln\left(1-\Gamma_\Delta\right) \right) \right]\,.
\end{equation}
It is non-negative, and vanishes only for $\Gamma=\Gamma_\Delta$. The latter property can be quantified in the form
\begin{equation}\label{in}
T\, \mathcal{H}(\Gamma,\Gamma_\Delta) \geq   {\rm Tr\,} \left[ K_T(H_\Delta) \left(\Gamma -\Gamma_\Delta\right)^2\right]\,.
\end{equation}

The identity (\ref{fi}) and the inequality (\ref{in}) hold true for
any choice of the function $\psi_0$ in (\ref{def:agl}). It is used
in two separate steps. One first simply takes $\psi_0=0$, and uses the
gap of the operator $K_{T_c}(H_0) + V$ to conclude that $\alpha$ is
close to an $\alpha_{\rm GL}$ for a suitable $\psi_0$, i.e., $\alpha$ is
of the form (\ref{def:agl}) to leading order. One then repeats the
argument with this choice of $\psi_0$, to conclude that the two terms
in the last line of (\ref{fi}) are negligible for small $h$.

To conclude the proof, it remains to calculate the terms in the first two lines in (\ref{fi}). For this purpose one needs semiclassical estimates with good regularity bounds. The relevant estimates can be summarized as follows:

\begin{theorem}
With errors controlled by $H^1$ and $H^2$ norms of $\psi_0$
\begin{align}\nonumber
  &-\frac {h^3} 2 T  \, {\rm Tr\,}  \left[ \ln\left(1 + e^{-H_\Delta/T}\right) -  \ln\left(1 + e^{-H_0/T} \right)\right] \\
  &= h^2 \mathcal{D}_2(\psi_0) +  h^4 \mathcal{D}_4(\psi_0) + \lambda_0 h^4
  \mathcal{E}^{\rm GL}(\psi_0) + O(h^5) \label{inn}
\end{align}￼￼
and
\begin{equation}\label{in2}
h^3\int V (|x - y|/h)|\alpha_{\rm GL}(x, y)|^2 \, dx\, dy  =h^2 \mathcal{D}_2(\psi_0) + h^4 \mathcal{D}_4(\psi_0)  +O(h^5)
\end{equation}
for suitable $\mathcal{D}_2$ and $\mathcal{D}_4$.
\end{theorem}

In (\ref{inn}) we have already taken into account the choice (\ref{def:D}) for the temperature $T$. For this choice, all the terms of order $h^2$ cancel, and one is left with $\lambda_0 h^4 \mathcal{E}^{\rm GL}(\psi_0)$ after taking the difference between (\ref{inn}) and (\ref{in2}).

This completes our very brief sketch of the proof of Theorem~\ref{main:thm}.

\section*{Acknowledgments}

Financial support via U.S. NSF grant PHY-1068285 (R.F.), NSERC (R.S.)
and a grant from the Danish council for independent research (J.P.S.)
is gratefully acknowledged.

\bibliographystyle{ws-procs975x65}
%\bibliography{ws-pro-sample}

\end{document}